%
\def\@{{\char'100}}

\long\def\abstract#1{\bigskip{\advance\leftskip by 2true cm
\advance\rightskip by 2true cm\eightpoint\centerline{\bf
Abstract}\everymath{\scriptstyle}\vskip10pt\vbox{#1}}\bigskip}
\long\def\resume#1{{\advance\leftskip by 2true cm
\advance\rightskip by 2true cm\eightpoint\centerline{\bf
R\'esum\'e}\everymath{\scriptstyle}\vskip10pt \vbox{#1}}}

\def\references{\bigbreak\centerline{\sc
References}\medskip\nobreak\bgroup
\def\ref##1&{\leavevmode\hangindent 15pt
\hbox to 15pt{\hss\bf[##1]\ }\ignorespaces}
\parindent=0pt
\everypar={\ref}\par}
\def\endreferences{\egroup}
\long\def\authoraddr#1{\medskip{\baselineskip9pt\let\\=\cr
\halign{\line{\hfil{\Addressfont##}\hfil}\crcr#1\crcr}}}
\def\Subtitle#1{\medbreak\noindent{\Subtitlefont#1.} }
%
%
\newif\ifrunningheads
\runningheadstrue
\immediate\write16{- Page headers}
\headline={\ifrunningheads\ifnum\pageno=1\hfil\else\ifodd\pageno\rightheadline
\else\leftheadline\fi\fi\else\hfil\fi}
\def\rightheadline{\sc\hfil\RightHeadText\hfil}
\def\leftheadline{\sc\hfil\LeftHeadText\hfil}

\hyphenation{Harnad Neumann}
%
%
\immediate\write16{- Fonts "Small Caps" and "EulerFraktur"}  
%
%
%

\let\sc=\tensmc
%
%
\font\teneuf=eufm10  \font\seveneuf=eufm7 \font\fiveeuf=eufm5
\newfam\euffam \def\gr{\fam\euffam\teneuf}

\textfont\euffam=\teneuf \scriptfont\euffam=\seveneuf 
\scriptscriptfont\euffam=\fiveeuf
%
\edef\smatrix[#1\&#2\\#3\&#4]{\left({#1 \atop #3}\, {#2 \atop #4}\right)}

\def \mt {\mapsto}
\def \ra {\rightarrow}

\def \a {\alpha}
\def \b {\beta}
\def \d {\delta}

\def \g {\gamma}
\def \G {\Gamma}

\def \l {\lambda}

\def \th {\vartheta}
\def \t {\tau}
\def \o {\omega}

\def \ss {\subset}

\def \mod{{\rm mod\,}}

\def\nchi{\hbox{\raise 2.5pt\hbox{$\chi$}}}
%
%

\def\grG{{\gr G}}

\def\nchi{\hbox{\raise 2.5pt\hbox{$\chi$}}}
%
%
\def\AA{{\cal A}}

%
%

		\def\bfC{{\bf C}}

		\def\bfH{{\bf H}}
		\def\bfI{{\bf I}}

		\def\bfP{{\bf P}}
		\def\bfQ{{\bf Q}}

		\def\bfZ{{\bf Z}}

%
%
\def\authorfont{\sc}
\font\eightrm=cmr8
\font\eightbf=cmbx8
\font\eightit=cmti8
\font\eightsl=cmsl8

\def\eightpoint{\let\rm=\eightrm \let\bf=\eightbf \let\it=\eightit
\let\sl=\eightsl \baselineskip = 9.5pt minus .75pt  \rm}

\font\titlefont=cmbx10 scaled\magstep2
\font\sectionfont=cmbx10
\font\Subtitlefont=cmbxsl10
\font\Addressfont=cmsl8
%
%
\def\Proclaim#1:#2\par{\smallbreak\noindent{\sc #1:\ }
{\sl #2}\par\smallbreak}
\def\Demo#1:#2\par{\smallbreak\noindent{\sl #1:\ }
{\rm #2}\par\smallbreak}
%
%
\immediate\write16{- Section headings}
\newcount\secount
\secount=0
\newcount\eqcount
\outer\def\section#1.#2\par{\global\eqcount=0\bigbreak
\ifcat#10
 \secount=#1\noindent{\sectionfont#1. #2}
\else
 \advance\secount by 1\noindent{\sectionfont\number\secount. #2}
\fi\par\nobreak\medskip} 
%
%
\immediate\write16{- Automatic numbering} 
\catcode`\@=11
\def\adv@nce{\global\advance\eqcount by 1}
\def\unadv@nce{\global\advance\eqcount by -1}
\def\nextnumber{\adv@nce}
%
%
\newif\iflines
\newif\ifm@resection
\def\onesec{\m@resectionfalse}
\def\moresec{\m@resectiontrue}
\moresec
\def\eq{\global\linesfalse\eq@}
\def\eqn{\global\linestrue&\eq@}
\def\nosubind@x{\global\subind@xfalse}
\def\newsubind@x{\ifsubind@x\unadv@nce\else\global\subind@xtrue\fi}
\newif\ifsubind@x
\def\eq@#1.#2.{\adv@nce
 \if\relax#2\relax
  \edef\loc@lnumber{\ifm@resection\number\secount.\fi
  \number\eqcount}
  \nosubind@x
 \else 
  \newsubind@x
  \edef\loc@lnumber{\ifm@resection\number\secount.\fi
  \number\eqcount#2}
 \fi
 \if\relax#1\relax
 \else 
  \expandafter\xdef\csname #1@\endcsname{{\rm(\loc@lnumber)}}
  \expandafter
  \gdef\csname #1\endcsname##1{\csname #1@\endcsname
  \ifcat##1a\relax\space
  \else
   \ifcat\noexpand##1\noexpand\relax\space
   \else
    \ifx##1$\space
    \else
     \if##1(\space
     \fi
    \fi
   \fi
  \fi##1}\relax
 \fi
 \eq@@{\loc@lnumber}}
\def\eq@@#1{\iflines \else \eqno\fi{\rm(#1)}}
\def\m@th{\mathsurround=0pt}
%
%
\def\display#1{\null\,\vcenter{\openup1\jot
\m@th
\ialign{\strut\hfil$\displaystyle{##}$\hfil\crcr#1\crcr}}
\,}
\newif\ifdt@p
\def\@lign{\tabskip=0pt\everycr={}}
\def\displ@y{\global\dt@ptrue \openup1 \jot \m@th
 \everycr{\noalign{\ifdt@p \global\dt@pfalse
  \vskip-\lineskiplimit \vskip\normallineskiplimit
  \else \penalty\interdisplaylinepenalty \fi}}}
%
%
\def\displayno#1{\displ@y \tabskip=\centering
 \halign to\displaywidth{\hfil$
\@lign\displaystyle{##}$\hfil\tabskip=\centering&
\hfil{$\@lign##$}\tabskip=0pt\crcr#1\crcr}}
%
%
\def\cite#1{{[#1]}}
\catcode`\@=\active
%
\hyphenation{}
%
\magnification=\magstep1
\hsize= 6.75 true in
\vsize= 8.75 true in 
%
%
\def\RightHeadText{Picard-Fuchs Equations, Hauptmoduls and Integrable Systems}
\def\LeftHeadText{J. Harnad}
%
%
\leftline{solv-int/9902013 \hfill CRM-2596 (1999) \break} \bigskip 
\bigskip \bigskip
\centerline{\titlefont Picard-Fuchs Equations, Hauptmoduls}
\centerline{\titlefont and Integrable Systems}
\bigskip
\centerline{\authorfont J.~Harnad}
\authoraddr
{Department of Mathematics and Statistics, Concordia University\\
7141 Sherbrooke W., Montr\'eal, Qu\'e., Canada H4B 1R6, {\rm \eightpoint
and} \\ 
Centre de recherches math\'ematiques, Universit\'e de Montr\'eal\\
C.~P.~6128, succ. centre ville, Montr\'eal, Qu\'e., Canada H3C 3J7\\
{\rm \eightpoint e-mail}: harnad\@crm.umontreal.ca} 
\bigskip

\abstract{The Schwarzian equations satisfied by certain Hauptmoduls
(i.e., uniformizing functions for Riemann surfaces of genus zero) are derived
from the Picard--Fuchs equations for families of elliptic curves and associated
surfaces.  The inhomogeneous Picard--Fuchs equations associated to elliptic
integrals with varying endpoints are derived and used to determine solutions
of equations that are algebraically related to a class of Painlev\'e VI
equations.} 
\bigskip \baselineskip 14 pt

\section 1. Differential equations for modular functions 

   There are a number of differential systems whose general solutions may be
expressed in terms of modular functions. An example is the Darboux-Halphen
system \cite{Ha},  
$$
\eqalign{
w_1' &=  w_1(w_2 + w_3) - w_2 w_3   \cr
w_2' &=  w_2(w_1 + w_3) - w_1 w_3    \cr
w_3' &=  w_3(w_1 + w_2) - w_1 w_2,}  \eq Halphen..
$$
which was already thoroughly studied in the last century, but recently has
recurred in several applications in mathematical physics \cite{GP, AH, CAC, T,
Hi, Du}. Its symmetrization under the symmetric group in three variables $S_3$
gives the Chazy equation \cite{C}
$$
W'''=2 W W'' - 3 W'^2, \eq Chazy..
$$
whose solutions are related to those of \Halphen by
$$
W=2(w_1+w_2+w_3). \eq Wdef..
$$
The physical contexts in which these equations have appeared include:
\item {1.}  The dynamics of magnetic monopoles pairs \cite{AH}.
\item {2.} Homogeneous, $SO(3)$ invariant solutions of self--dual
Einstein equations \cite{GP, T, Hi}. 
\item {3.}  Solutions of the WDVV equations in topological field theory
\cite{Du}.

  The general solution to \Halphen was determined in 1881 by Halphen \cite{Ha}
and Brioschi \cite{Br} in terms of the {\it elliptic modular function}
$$
\lambda(\tau) = k^2(\tau),  \eq ellipticmodular..
$$
where $k(\tau)$ is the elliptic modulus, viewed as a function of the ratio $\tau$
of the elliptic periods. A particular solution to \Halphen is given by
$$
\eqalign{
w_1 &:=  {1\over 2}{d\over d\t}\ln {\l'\over \l},\quad
w_2 :=  {1\over 2}{d\over d\t}\ln {\l'\over (\l-1)},\cr
w_3 &:=  {1\over 2}{d\over d\t}\ln {\l'\over \l(\l-1)}.} \eq Halphensol..
$$
The general solution is obtained by composing this with a general M\"obius
transformation
$$
T:\t \ra {a\t  + b \over c \t + d} \equiv T(\t), 
\qquad \pmatrix{a & b \cr c & d } \in  SL(2, \bfC).  \eq Mobius..
$$

 Similarly, a particular solution to \Chazy is given by
$$
W := 2\sigma_1 ={1\over 2}{d\over d\t} \ln{J'^6\over J^4(J-1)^3}, \eq
Chazysol.. 
$$
 where $J$ denotes Klein's $J$--function
$$
J = {4(\l^2 -\l +1)^3\over 27 \l^2 (\l-1)^2},  \eq JKlein..
$$
and the general solution is again obtained by composing $J$ with a M\"obius
transformation \Mobius.

   The key element in deriving these solutions is to first note that the period
integrals,  when viewed as functions of $\l$, are hypergeometric functions
$$
\eqalign{
K_1 &= \int_0^1 {dt \over \sqrt{(1-t^2)(1- \lambda t^2)}} 
= {\pi\over 2} F({\scriptstyle{1\over 2},  {1\over 2}}; 1; \lambda) 
 \cr  iK_2 &= \int_0^1 {dt 
\over \sqrt{(1-t^2)(1- (1-\lambda) t^2)}}  = {i\pi \over 2} 
 F({\scriptstyle{1\over 2},
{1\over 2}}; 1;1- \lambda) ,}  \eq ellipticintegrals..
$$
and hence satisfy the hypergeometric equation of Legendre type
$$
\l(1-\l) {d^2 y \over d \l^2} + (1-2\l) {dy\over d\l} - {1\over 4} y = 0. 
\eq Legendrehypergeom..
$$
Since $\l(\tau)$ is the inverse of the function $\tau(\l)$ given by the ratio
of the periods, it follows that it satisfies the Schwarzian equation
$$
\{\l, \t\} + {\l^2 - \l + 1\over 2\l ^2(1 -\l)^2 }\l'^2 = 0,   
\eq lambdaSchwarzian..
$$
where 
$$
\{f, \t\} := {f'''\over f'} -{3\over 2} \left({f''\over f'}\right)^2, 
\qquad  (f':={df\over d\t}). \eq Schwarzian..
$$
is the Schwarzian derivative \cite{GS}. This implies that \Halphensol defines a
solution of \Halphen and, by the $SL(2, \bfC)$ invariance of the system,
composition with the M\"obius transformations \Mobius gives the general
solution. Similarly, the fact that \Chazysol determines the general solution to
\Chazy follows from the Schwarzian equation satisfied by $J$
$$
\{J, \t\}+ { 36J^2 - 41J + 32\over 72 J^2(J-1)^2} J'^2 = 0, \eq JSchwarzian..
$$
which is obtained from \lambdaSchwarzian by composing with \JKlein. This
latter Schwarzian equation equation is analogously related to the  hypergeometric
equation
$$
J(1-J) {d^2y \over dJ^2} + \left({2\over 3} - {7\over 6} J\right) {dy\over dJ}
-{1\over 144} y = 0.  \eq Jhypergeom.. 
$$
Thus $J$ is the inverse of the function obtained taking  the ratio of two
linearly independent solutions of \Jhypergeom.

  The hypergeometric equations \Legendrehypergeom and \Jhypergeom may both be
viewed as examples of {\it Picard--Fuchs} equations for elliptic pencils; that
is, as Fuchsian differential equations determining the variation of elliptic
integrals over an affine parametric family of elliptic curves. In each of these
cases, the relevant inverse function is a modular function. This follows from
the fact that the projectivized monodromy groups for the associated
hypergeometric equations \Legendrehypergeom and \Jhypergeom are both
commensurable with the modular group $\Gamma :=PSL(2,\bfZ)$. For the case
\Legendrehypergeom, this gives the principal congruence subgroup $\Gamma(2)$,
which is the automorphism group of the modular functions $\l(\tau)$, while for
\Jhypergeom, it is the full modular group, the automorphism group of $J(\tau)$. 

   There is another sense in which Picard--Fuchs equations may be related to
nonlinear equations of interest in mathematical physics; namely, the class of
isomonodromic deformation equations, such as the family of Painlev\'e equations
$P_{VI}(\a,\b,\g,\d )$
$$
\eqalign{
X'' =&{1\over 2} \left({1\over X} + {1\over X-1} + {1\over X-t}\right)
(X')^2  - \left({1\over t} + {1\over t-1} +  {1\over X-t}\right)X' 
\cr
&\quad +{X(X
-1)(X-t)\over t^2(t-1)^2}\left(\a + {\b t^2\over X^2}+ {\g(t-1)\over (X-1)^2}
+{ \d t(t-1)\over (X-t)^2}\right).} \eq PainleveVI..
$$
It was shown in the work of Hitchin \cite{Hi}, Dubrovin \cite{Du} and Mazzocco
\cite{M} that solutions  to particular cases of \PainleveVI with special
values of the parameters $(\a,\b,\g,\d)$ could be given in terms of 
solutions to  \Halphen. For example, for the case $P_{VI}(\a=2, \b=0, \g=0,
\d={1\over 2})$, we have the one--parameter family of Chazy solutions given
by 
$$
X ={(w_2 w_3 - w_1 w_2 - w_1 w_3)^2\over 4w_1 w_2 w_3 (w_1 - w_3)}, 
\eq ChazyPVI..
$$
where the independent parameter is taken as
$$
t:= {w_1-w_3\over w_2 - w_3} = \l, \eq..
$$
and $(w_1,w_2,w_3)$ is a general solution to \Halphen. (Only one of the
three $SL(2, \bfC)$ parameters introduced by \Mobius is effective.)  More
generally, a sequence of Chazy--type solutions for parameter values  $\a={1\over
2}(2\mu -1)^2,\mu +{1\over 2} \in \bfZ \backslash 1$ are derived in \cite{M} by
application of discrete symmetry transformations.

  Another class of solutions to \PainleveVI for the case $(\a=0, \b=0, \g=0,
\d={1\over 2})$ was already known to Picard \cite{Pi}, who expressed
them in terms of elliptic integrals with variable end-points. This allows us to
relate this case to an {\it inhomogeneous } Picard--Fuchs equation.
Namely, consider the $1$--parameter family of elliptic curves
$$
y^2 = 4x(x-1)(x-\l), \eq..
$$
 and corresponding period integrals:
$$
K_1= \oint_\infty^1 {dx\over y}, \qquad  K_2= \oint_\infty^0 {dx\over y}. \eq..
$$
If the elliptic integral with {\it varying} ($\l$--dependent) endpoints 
$$
K := \int_{\infty}^{(X(\l),Y(\l))}{dx\over y} \eq..
$$
is required to  satisfy the same hypergeometric equation as do the period
integrals $K_1,\ K_2$, i.e., if it is set equal to a linear combination
$$
K = A K_1 + B K_2,  \eq..
$$
it follows that  $X=X(\l)$ satisfies $P_{VI}(\a=0,\b=0,\g=0,\d={1\over
2})$, providing a $2$--parameter family of solutions. More generally,  R. Fuchs
(1907) in \cite{Fu} showed that 
$P_{VI}(\a,\b,\g,\d)$ is equivalent to the inhomogeneous Picard--Fuchs
equation
$$
\eqalign{
\l(1-\l){d^2 K \over d\l^2}& + (1-2\l){d K\over d\l}
-{1\over 16} K \cr
& ={Y\over \l(\l-1)}
\left(\a + {\b\l^2\over X^2}+
{\g(\l-1)\over (X-1)^2}
+\Big(\d-{1\over 2}\Big){\l(1-\l)\over (X-\l)^2}\right) .}   \eq InhomPF..
$$
 (A more recent perspective on such equations and their algebro--geometric
meaning may be found in \cite{Ma}.)

   In the following sections, a number of further examples of modular functions
having similar properties will be considered. These all provide solutions to
certain associated systems of nonlinear differential equations whose origins
may be traced to Picard--Fuchs equations for families of elliptic curves. In each
case, an associated inhomogeneous Picard--Fuchs equation may also be
determined, and shown equivalent to an equation that is algebraically related
to the Picard case of $P_{VI}$.

\section 2. Generalized Halphen Equations

\Subtitle {2a. Triangular cases}
\smallskip
\nobreak
 
  Halphen \cite{Ha} also considered generalizations of the system
\Halphen related to the general hypergeometric equation 
$$
f(1-f) {d^2y \over df^2} + (c - (a + b + 1) f) {d y \over df} - a b y = 0.
\eq hypergeometric..
$$
Assuming  the inverse function of the ratio of two linearly independent
solutions of \hypergeometric to exist (which  by no means is always the
case in a global sense, in view of the infinite-valued multiplicity of
the solutions of \hypergeometric due to monodromy), it also satisfies a
Schwarzian equation of the form
$$
\{f, \t\} + 2R(f) f'^2=0, \eq Schwarzianeqabc..
$$
where
$$
R(f) = {1\over 4}\left({1-\l^2\over f^2} + {1-\mu^2\over (f-1)^2} + 
{\l^2 +\mu^2 -\nu^2 -1 \over f(f-1)}\right),  \eq rationalabc..
$$
with the parameters $(\l, \mu, \nu)$ defined by
$$
\l := 1-c, \quad \mu := c-a-b, \quad \nu := b-a . \eq..
$$
The generalized Halphen-Brioschi variables
$$
\eqalign{
W_1 &:= {1\over 2}{d\over d\t}\ln {f'\over f},\quad
W_2 := {1\over 2}{d\over d\t} \ln {f'\over (f-1)},\cr
W_3 &:= {1\over 2} {d\over d\t}\ln {f'\over f(f-1)}} 
\eq Wgeneraldef.. 
$$
then satisfy the {\it general} Halphen system:
$$
\eqalign{
W_1' &= W_1( W_2+W_3) - W_2 W_3 + X \cr
W_2' &= W_2( W_1+W_3) - W_1 W_3 + X \cr
W_3' &= W_3( W_1+W_2) - W_1 W_2 + X, \cr
X &:= \mu^2 W_1^2 + \l^2 W_2^2 + \nu^2 W_3^2 +(\nu^2 - \l^2 -\mu^2)W_1 W_2 \cr
&+(\l^2 - \mu^2 -\nu^2)W_3 W_1+(\mu^2 - \l^2 -\nu^2)W_2 W_3.} 
\eq Wgeneraleq.. 
$$

   A sufficient condition for having a well-defined inverse function $f(\tau)$
is that the projectivized monodromy group of the hypergeometric equation
\hypergeometric be a Fuchsian group of the first kind. This essentially means
that it acts properly discontinuously and there is a tesselation of the
image space by fundamental domains with a finite number of vertices. In the
present case, the domains are necessarily triangular (with circular arcs as
sides), since the vertices must map to the three regular singular points
$(0,1,\infty)$. (The more general case, with $n$ singular points is
considered in the next subsection.) The projectivized image of the monodromy
representation is just the automorphism group of the function. The parameters
$(\l, \mu, \nu)$ are the fractions of $\pi$ giving the angles at the vertices of
the fundamental domain. 

   In general, the automorphism group $\grG_f$ of a function $f(\tau)$ (under
M\"obius transformations \Mobius) is said to be {\it commensurable} with the
modular group $\G$ if it is a subgroup of $PGL(2,\bfQ)$ whose intersection
with $\Gamma$ is of finite index in both $\Gamma$ and $\grG_f$. Such a function
and its automorphism group will be referred to as {\it modular}.  By a {\it
Hauptmodul}, we understand a  uniformizing function for a genus zero Riemann
surface; that is, the quotient $\bfH/\grG_f$ of the upper half $\tau$--plane by
the automorphism group defines a genus zero Riemann surface. The
fact that such a function is the sole generator of the field of meromorphic
functions on the Riemann surface implies that $f$ must satisfy a Schwarzian
equation of the type \Schwarzianeqabc for {\it some} rational function $R(f)$. 
A special class of such Hauptmoduls, referred to as {\it replicable functions} 
(due to their replication properties under the action of generalized Hecke
operators
\cite{CN}), have the additional property of containing a finite index subgroup
of the type
$$
\G_0(N) :=\left\{\pmatrix {a & b \cr c & d } \in  SL(2,\bfZ), \ c\equiv 0 \
\mod N\right\}. \eq Gammazeron..
$$
(Such functions arise in connection with ``modular moonshine'', either as
character generators for the Monster sporadic group, or in relation to these
through the generalized Hecke averaging procedure \cite{CN, FMN}). Each such
function, of which there are only a finite number, is analytic in the upper
half--plane, has a finite number of vertices in its fundamental domains, a cusp
at $\infty$ and admits, up to an affine transformation, an expansion as a
normalized McKay--Thompson
$q$--series
$$
F(q) = \a f(\tau) + \b ={1\over q} + \sum_{n=1}^\infty a_n q^n, 
 \qquad q:= e^{2i\pi \t}, \eq..
$$
convergent in the upper half--plane. (For the cases considered here, we also
have $ a_n\in \bfZ$.)

  The table below (which is taken from \cite{HM}) contains a complete list, up to
equivalence under affine transformations in the $\tau$ variable, of the
replicable functions of triangular type (i.e., those for which the fundamental
domain has three vertices). These coincide with the arithmetic triangular
functions of noncompact type \cite{Ta}. Through formulae \Wgeneraldef, they
provide solutions to the general Halphen systems \Wgeneraleq.
\medskip

\centerline{\bf{Table 1.  Triangular Replicable Functions}}\nobreak
\nobreak
\bigskip
\centerline{\hskip-20pt
\vbox{\tabskip=0pt \offinterlineskip
\def\tablerule{\noalign{\hrule}}
\halign to330pt{\strut#& \vrule#\tabskip=.5em plus2em&
 \hfil#\hfil & \vrule # &\hfil #\hfil & \vrule # &
 \hfil#\hfil & \vrule# & \hfil#\hfil & \vrule# &
 \hfil#\hfil & \vrule# & \hfil#\hfil & \vrule#
\tabskip=0pt\cr\tablerule
&& Name && $(a,b,c)$ && $(\lambda, \mu, \nu)$ && $\rho_0$ 
&& $f(\tau)$   &\cr \tablerule
&& $\matrix{1A \cr \sim \G}$ && $({1\over 12},{1\over 12},{2\over 3})$ 
&& $({1\over 3},{1\over 2},0)$  &&
$\smatrix[ 0 \&-1 \\ 1 \& -1] $
&& $J = {(\vartheta_2^8 + \vartheta_3^8 + 
\vartheta_4^8)^3\over  54 \vartheta_2^8 \vartheta_3^8 \vartheta_4^8}$
&\cr \tablerule
&& $2A$ && $({1\over 8}, {1\over 8}, {3\over 4})$ 
&& $({1\over 4}, {1\over 2}, 0)$ &&
$\smatrix[ 0 \&-1 \\ 2 \& -2] $ 
&&  
${ \left(\vartheta_3^4 +\vartheta_4^4\right)^4 \over
16  \vartheta_2^8 \vartheta_3^4 \vartheta_4^4}$
&\cr\tablerule 
&& $3A$ && $({1\over 6}, {1\over 6}, {5\over 6})$ && 
$({1\over 6}, {1\over 2}, 0)$  && 
$\smatrix[ 0 \&-1 \\ 3 \& -3] $
&&  ${\left(\eta^{12}(\tau) + 27 \eta^{12}(3\tau)\right)^2
\over 108 \eta^{12}(\tau)\eta^{12}(3\tau)}$ 
&\cr\tablerule
&& $\matrix{2B \cr \sim\G_0(2)}$ && $({1\over 4}, {1\over 4}, {1\over 2})$ && 
$({1\over 2}, 0, 0)$  && 
$\smatrix[ 1 \&-1 \\ 2 \& -1]$
&&  $\matrix{1 + {1\over 64}\big({\eta(\tau)\over
\eta(2\tau)}\big)^{\scriptscriptstyle 24}\cr = {\big(\vartheta_3^4(\tau) +
\vartheta_4^4(\tau)\big)^2\over
\vartheta_2^8(\tau)}} $
&\cr\tablerule
&& $\matrix{3B \cr \sim \G_0(3)}$ && $({1\over 3}, {1\over 3}, {2\over 3})$ && 
$({1\over 3}, 0, 0)$  && 
$\smatrix[ 1 \&-1 \\ 3 \& -2] $ 
&&  $1 + {1\over 27}\big({\eta(\tau)\over
\eta(3\tau)}\big)^{\scriptscriptstyle 12}$ 
&\cr\tablerule
&& $\matrix{4C^*\cr \sim \G_0(4)}$ && $({1\over 2}, {1\over 2}, 1)$ && 
$(0, 0, 0)$  && 
$\smatrix[ 1 \&-1 \\ 4 \& -3] $ 
&&  $\matrix{{1\over \l(2\tau)}
 = {\vartheta_3^4(2\tau) \over \vartheta_2^4(2\tau)}\cr 
=1 +{1\over 16}\big({\eta(\tau)\over\eta(4\tau)}\big)^{\scriptscriptstyle 8}
\cr }$ &\cr\tablerule
&& $2a$ && $({1\over 6}, {1\over 6},  {2\over 3})$ && 
$({1\over 3}, {1\over 3}, 0)$  && 
$\smatrix[ 2 \&-3 \\ 4 \& -4] $
&&  
${\sqrt{3}i\left( e^{\pi i/3}\th_3^4(2\tau) -\th_2^4(2\tau)\right)^3\over
9\th_2^4(2\tau)\th_3^4(2\tau)\th_4^4(2\tau)}$
&\cr\tablerule
&& $4a$ && $({1\over 4}, {1\over 4}, {3\over 4})$ && 
$({1\over 4}, {1\over 4}, 0)$  && 
$\smatrix[ 4 \&-5\\ 8 \& -8]$ 
&& $-{i\left(\th_3^2(2\tau) +
i \th_4^2(2\tau)\right)^4\over 8\th_2^4(2\tau)\th_3^2(2\tau)\th_4^2(2\tau)}$
&\cr\tablerule
&& $6a$ && $({1\over 3}, {1\over 3}, {5\over 6})$ && 
$({1\over 6}, {1\over 6}, 0)$  && 
$\smatrix[ 6 \&-7 \\ 12 \& -12]$
&&  
$-{\sqrt{3}i \left(\eta^6(2\tau) + 3\sqrt{3}i\eta^6(6\tau)\right)^2\over
36\eta^6(2\tau)\eta^6(6\tau)}$
&\cr\tablerule 
\hfil\cr}}}

In this table, the first column gives the labelling according to the notation
of refs.~\cite{CN, FMN}, consistent with the finite group atlas. The
second and third columns give the hypergeometric parameters $(a,b,c)$
and $(\lambda,\mu,\nu)$. The fourth column contains the  generator of the
automorphism group which stabilizes a vertex mapping to $0$. The
corresponding generator stabilizing the cusp at $\infty$ is 
$$
\rho_\infty\ = \smatrix[ 1 \& 1 \\ 0 \& 1], \eq..
$$ 
and the third generator $\rho_1$, stabilizing a vertex mapping to $1$
is determined by the relation
$$
\rho_\infty \rho_1 \rho_0 = \bfI.  \eq..
$$
The last column in Table 1 gives explicit expressions for the Hauptmoduls in
terms of null theta functions $\vartheta_2(\tau), \vartheta_3(\tau),
\vartheta_4(\tau)$ or the Dedekind eta--function $\eta(\tau)$. 

   In section 3, it will be shown how the corresponding hypergeometric
equations,  which imply the Schwarzian equation \Schwarzianeqabc, and hence
provide solutions  to the generalized Halphen equations for the triangular
cases, may be derived as Picard--Fuchs equations for families of elliptic
curves. But first we consider some further generalizations of the Halphen system
\Halphen corresponding to second order Fuchsian equations with more than three
regular singular points. Such generalized systems were introduced by Ohyama
\cite{Oh}.

\Subtitle {2b. Generalized Halphen Systems with $n$ singular points}
\smallskip
\nobreak
   Consider second order Fuchsian equations of the form
$$
{d^2 y \over df^2} + R(f) y = 0,  \eq fuchseqgen..
$$
where $R(f)$ is a rational function of the form 
$$
R(f) ={N(f)\over (D(f))^2}, \qquad
D(f) = \prod_{i=1}^n (f-a_i),  \eq..
$$
and $N(f)$ is a polynomial of degree $\leq 2n - 2$. (Any second order Fuchsian
equation is projectively equivalent to one of this form; i.e., it may be
transformed to this form, with no first derivative term, by multiplication of
the solutions by a suitably chosen function.) Let $\tau(f)$ again denote the
ratio
$$
\tau(f):={y_1\over y_2}  \eq taugen..
$$
of two linearly independent solutions of  \fuchseqgen, and suppose again that
the inverse function $f=f(\tau)$ is well-defined. It then satisfies the
Schwarzian differential equation
$$
\{f, \t\}+ 2 R(f) f'^2  = 0,  \eq Schwarzeqgen..
$$
and conversely, at least locally, all solutions of the Fuchsian equation 
\fuchseqgen are expressible as:
$$
y = {(A + B\t (f)) \over (\t')^{1\over 2}}.   \eq..
$$

The image of the monodromy representation
$$
\eqalign{
M: \pi_1(\bfP -\{a_1, \dots a_n, \infty\}) &\ra GL(2,\bfC) \cr
M: \g & \mt M_\g =: \pmatrix {a & b \cr c & d},} \eq..
$$
defined up to global conjugation by
$$
\g : (y_1, y_2)\vert_{f_0} = (y_1, y_2)\vert_{f_0} M_\g,  \eq..
$$
determines a subgroup $\grG_f \ss GL(2, \bfC)$ that acts on $\tau$
by M\"obius transformations \Mobius, leaving $f(\tau)$ invariant.  
Introducing the new variables \cite{Oh, HM}
$$
u:= X_0 = {1\over 2} {f'' \over f'}, \quad v_i :={1\over 2}( X_0 - X_i) =
{1\over 2}{f'\over f- a_i}, \eq..
$$
these satisfy the set of quadratic constraints 
$$
(a_i - a_j) v_i v_j + (a_j - a_k) v_j v_k
+ (a_k - a_i) v_k v_i  = 0,   \eq quadconstr..     
$$
and the differential equations:
$$
\eqalignno{
v'_i&= -2 v_i^2 + 2 u v_i, \qquad i=1,\dots n  \eqn diffeqv.a. \cr
u' &= u^2 - \sum_{i,j=1}^n r_{ij} v_i v_j, \eqn diffequ.b. }
$$
where the quadratic form $\sum_{i,j=1}^n r_{ij} v_i v_j$ appearing \diffequ  is
determined by expressing of the rational function $R(f)$ in the form
$$
R(f) = {1\over 4}\sum_{i,j=1}^n {r_{ij}\over (f-a_i)(f- a_j)}.  \eq Rfromr..
$$
(There is a nonuniqueness in such expressions for $R(f)$, but this just
corresponds to the freedom of adding any linear combination of the
vanishing quadratic forms \quadconstr to the right hand side of eq.~\diffequ.) 
In this case, if the automorphism group of $f$ is again Fuchsian, the angles
$\{\a_i \pi\}_{i=1, n}$ at the finite vertices of the fundamental polygon are
related to the diagonal part of the quadratic form by 
$$
r_{ii} = 1 -\a_i^2. \eq..
$$

Table 2 below, which is a shortened version of one appearing in \cite{HM}, again
contains a list of Hauptmoduls that are replicable functions. But in these
cases, their fundamental domains have four vertices, and hence there are four
generalized Halphen variables $(u, v_1, v_2, v_3)$ appearing in eqs.~\diffeqv,
\diffequ.

The first column again identifies the functions and their groups according to
the notation of refs.~\cite{CN, FMN}, the second gives the values of $f(\tau)$ at
the finite vertices (i.e., the location of the poles of $R(f)$), and the third
lists the generators of the automorphism group of $f$ corresponding to these 
vertices (i.e., the projectivized monodromy group generators). The fourth column
gives the quadratic form appearing in \diffequ, and serves to define
the rational function $R(f)$ through \Rfromr. The last column lists explicit
formulae for $f(\tau)$ in terms of the Dedekind $\eta$--function. This table
contains all the geometrical and group theoretical data characterizing the
Hauptmoduls listed. A similar set of data may be determined for all the
replicable functions appearing in \cite{CN, FMN}, and from this the
corresponding Schwarzian equations and generalized Halphen equations may be
deduced. (It should be noted that the number of vertices for the corresponding
fundamental domains never exceeds $26$, and that there are algebraic relations
interlinking all these various cases.)

\centerline{\bf{Table 2.  Four Vertex Replicable Functions}}
\nobreak
\bigskip
\centerline{\hskip-20pt
\vbox{\tabskip=0pt \offinterlineskip
\def\tablerule{\noalign{\hrule}}
\halign to327pt{\strut#& \vrule#\tabskip=.5em plus1em&
 \hfil#\hfil & \vrule # &\hfil #\hfil & \vrule # &
 \hfil#\hfil & \vrule# & \hfil#\hfil & \vrule# &
 \hfil#\hfil & \vrule# & \hfil#\hfil & \vrule#
\tabskip=0pt\cr\tablerule
&& Name && $(a_1,a_2,a_3)$ && $\matrix{\rho_1 \cr \rho_2 \cr \rho_3}$ 
&& $\displaystyle{ \sum_{i,j=1}^3 r_{ij}v_i v_j} $ &&$f(\tau)-1$  
&\cr \tablerule
&& $\matrix{6C}$ && $(-3,0,1)$ && $\matrix{\smatrix[3 \& -2\\ 6 \& -3]\cr 
\smatrix[3 \& -1 \\ 12 \& -3]\cr 
\smatrix[-1 \& \phantom{-}0\\ \phantom{-}6 \& -1]}$  
&& $\matrix{{3\over 4}v_1^2 + {3\over 4} v_2^2 + v_3^2\cr
 - {1\over 2} v_2 v_3 - v_1 v_3}$ 
&& ${1\over 4} {\eta^6(\t) \eta^6(3\t) \over \eta^6(2\t)
\eta^6(6\t)}$ 
&\cr\tablerule
&& $\matrix{6D}$ && $\matrix{(\b,\bar{\b},1) \cr
\b:=\cr-{3\over 4} + \sqrt{2}i}$ 
&&
$\matrix{ \smatrix[4 \& -3 \\ 6 \& -4]\cr
\smatrix[2 \& -1\\ 6 \& -2]\cr \smatrix[-1 \& \phantom{-}0\\
\phantom{-}6 \& -1]}$  && $\matrix{{3\over 4}v_1^2 + {3\over 4} v_2^2 + v_3^2
\cr
 + {131\over 162} v_1 v_2 \cr - {28 -16\sqrt{2}i\over 81}v_1 v_3 \cr
- {28 +16\sqrt{2}i\over 81}v_2 v_3}$ 
&& ${1\over 4} {\eta^4(\t) \eta^4(2\t) \over \eta^4(3\t)
\eta^4(6\t)}$ 
&\cr\tablerule
&& $\matrix{6E\cr (\G_0(6))}$ && $(-{1\over 8},0,1)$
 && $\matrix{\smatrix[5 \& -3\\ 12 \& -7]\cr 
\smatrix[5 \& -2 \\ 18 \& -7]\cr 
\smatrix[-1 \& \phantom{-}0\\ \phantom{-} 6 \& -1]}$  
&& $\matrix{v_1^2 + v_2^2 + v_3^2\cr
 - {10\over 9} v_2 v_3 -{8\over 9} v_1 v_3}$ 
&& ${1\over 8} {\eta^5(\t) \eta(3\t) \over \eta (2\t)
\eta^5(6\t)}$ 
&\cr\tablerule
&& $\matrix{8E\cr ( \G_0(8))}$ && $(-1,0,1)$ 
&& $\matrix{\smatrix[3 \& -2\\ 8 \& -5]\cr 
\smatrix[3 \& -1 \\ 16 \& -5]\cr 
\smatrix[-1 \& \phantom{-}0 \\ \phantom{-}8 \& -1]}$  
&& $\matrix{v_1^2 +  v_2^2 + v_3^2\cr - 2v_1 v_3 }$ 
&& $\matrix{{1\over 4} {\eta^4(\t) \eta^2(4\t) \over
\eta^2(2\t) \eta^4(8\t)}\cr}$ 
 &\cr\tablerule
&& $\matrix{9B\cr (\G_0(9))}$ 
&& $\matrix{(\o,\bar{\o},1)\cr \o:= e^{2\pi i\over 3}}$ &&
$\matrix{ \smatrix[5 \& -4 \\ 9 \& -7]\cr \smatrix[2 \& -1\\ 9 \& -4]\cr 
\smatrix[-1 \& \phantom{-}0\\ \phantom{-}9 \&-1]}$  
&& $\matrix{v_1^2 +  v_2^2 + v_3^2\cr
 - v_1 v_2 \cr - (1-\o)v_1 v_3 \cr - (1-\bar{\o})v_2 v_3}$ 
&& ${1\over 3} {\eta^3(\t) \over \eta^3(9\t)}$ 
&\cr\tablerule
\hfil\cr}}}

\section 3. Picard--Fuchs Equations on elliptic families

 In this section, the parametric families of elliptic curves whose
associated Picard--Fuchs equations underlie the Scwharzian equations governing
these Hauptmoduls will be given for three of the examples appearing in the
tables above. Only those cases are treated which are actually subgroups of
the full modular group $\Gamma$, but an indication will be given at the
end of this section how the other cases may be similarly derived. (A
more complete version of these results will appear elsewhere \cite{H}.)   

\Subtitle {3a. Arithmetic triangular subgroups of \ $\Gamma$}
\smallskip
\nobreak
 
  Four of the cases appearing in Table 1  involve automorphism groups that
are contained in the full modular group; these are: $1A\sim \Gamma$, $2B\sim
\Gamma_0(2)$, $3B\sim \Gamma_0(3)$ and $4C\sim \Gamma_0(4) \sim \Gamma(2)$.
For each of these, it is possible to find a $1$--parameter family of elliptic
curves for which the associated Picard--Fuchs equation gives the required
hypergeometric equation. We illustrate this below for the two cases: $1A\sim
\Gamma$ and $2B\sim \Gamma_0(2)$. The first of these leads to the
hypergeometric equation \Jhypergeom associated with the $J$--function. (The
case $4C\sim \Gamma_0(4) \sim \Gamma(2)$ gives the corresponding Legendre
hypergeometric equation \Legendrehypergeom associated with $\l$, and hence the
original Halphen system.)

\noindent  1. Hauptmodul $1A$. The associated family of elliptic
curves  is given by the elliptic pencil 
$$
y^2 =4(a-1)x^3 - 3 a x -a \qquad (a=J).  \eq..
$$
Denote by  $K$ and $L$ the elliptic integrals of the first and
second kind,
$$
K:= \int_\infty^{(X,Y)} {dx\over y}, \qquad  L:= \int_\infty^{(X,Y)} {x\,dx\over
y},
\eq ellipticKL..
$$
where we allow the endpoints to depend upon the parameter $a$.
Differentiating these with respect to $a$,  taking the
endpoint contributions into account, we deduce the {\it inhomogeneous}
Gauss-Manin system
$$\eqalignno{
K' +{(5a-2)K\over 12a(a-1)}-{L\over 6a} &=
{a + (2+a)X + 2(1-a)X^2\over 6a(a-1)Y} + {X'\over Y} \eqn GammaInhomGMK.a. \cr
L' +{K+(14a+4)L\over 24a(a-1)} &=
{-a - aX + (4+2a)X^2\over 12a(a-1)Y} + {X X'\over Y}. \eqn GammaInhomGML.b.}
$$
(Usually, the term  ``Gauss--Manin system'' refers to the equations satisfied by
the corresponding differential cohomology classes,  but here we consider 
integrals along a suitable path, with varying endpoints.) If instead of taking 
variable endpoints, we integrate around a cycle, the right hand sides of
eqs.~\GammaInhomGMK, \GammaInhomGML vanishes, and we have the more usual
homogeneous Gauss--Manin system.  Eliminating the $L$ integral from this system
gives the  inhomogeneous Picard--Fuchs equation
$$
K''+{(2a-1)\over a(a-1)}K' +{(36a^2-41a-4)\over 144a^2(a-1)^2}K= 
{1\over Y} \left( X'' -\AA_2(X)X'^2-\AA_1(X)X' -\AA_0(X)\right), 
\eq GammaInhomPF..
$$
where
$$
\eqalignno{
\AA_2(X)&:={12(a-1)X^2-3a\over 2Y^2}   \eqn GammaAAtwo.a.\cr
\AA_1(X)&:={a^2 + 3a^2X - 4(a-1)^2X^3\over
    a(a-1)Y^2} \eqn GammaAAone.b.\cr
\AA_0(X)&:={1\over 72a^2(a-1)^2Y^2}\Big(-3a^2 + 
(16a - 34a^2)X \cr
&\quad+(60a - 87a^2)X^2 +(16 - 140a + 124a^2)X^4\Big)\eqn GammaAAzero.c.\cr
Y^2&:=4(a-1)X^3-3aX-a. \eqn GammaY.d.} 
$$

Taking the path defining $K$ to be a cycle, the right hand side of
\GammaInhomPF disappears and we obtain the homogeneous Picard--Fuchs equation,
which is projectively equivalent to the hypergeometric equation \Jhypergeom
under the identification $a=J$. (More  precisely, the function $a^{1\over
6}(a-1)^{1\over 4}K$, taken over a generating pair of cycles gives a basis of
solutions of \Jhypergeom.) Setting the integral $K$ with variable endpoints
equal to a linear combination 
$$
K= A K_1 + B K_2, \eq lincombK..
$$
where $K_1$ and $K_2$ denote the values of the integral taken over a basis of
cycles, amounts to choosing $X(a)$ as an elliptic function, defined as the
inverse of the elliptic integral $K$, with its argument taken as 
$A K_1 + B K_2$. This implies that the right hand side of  of \GammaInhomPF
disappears, giving an equation for $X(a)$ of the same type as Picard's case of
$P_{VI}$. (In fact, the two are algebraically related through through \JKlein.)

\noindent 2. Hauptmodul 2B. The associated family of elliptic
curves  in this case is given by the elliptic pencil 
$$
y^2 =4x^3 - 3(1+3 a)x+9a-1.  \eq ellipticfamilytwoB..
$$
(To be precise, the Hauptmodul $f_{2B}$ normalized as in Table 1 corresponds to
the inverse $1\over a$ of the parameter appearing in \ellipticfamilytwoB.)
 Denoting again by  $K$ and $L$ the elliptic integrals of the first and second
kind, respectively, defined as in \ellipticKL, and differentiating with respect
to the parameter, we obtain the corresponding inhomogeneous Gauss-Manin system:
$$
\eqalignno{
K' +{(3a-1)K-2L\over 12a(a-1)} &=
{1 +3a + (1-3a)X - 2X^2\over 6a(a-1)Y} + {X'\over Y} \eqn GammazerotwoGMK.a.\cr
L' +{(1+3a)K+(2-6a)L\over 24a(a-1)} 
&= -{1-9a + (1+3a)X - (2-6a)X^2\over
12a(a-1)Y} + {X X'\over Y} \eqn GammazerotwoGML.b.}
$$
and the resulting inhomogeneous Picard--Fuchs equation of type $2B$
$$
\eqalign{
K''+{(2a-1)\over a(a-1)}K' +{3\over 16a(a-1)}K= 
{1\over Y} \left( X'' -\AA_2(X)X'^2-\AA_1(X)X' -\AA_0(X)\right)}, 
\eq GammazeroInhomPF..
$$
where
$$
\eqalignno{
\AA_2(X)&:={1\over 2(X-1)}+{2+4X\over 1-9a+4X+4X^2}  \eqn GammazeroAAtwo.a.\cr
\AA_1(X)&:=-{1\over a-1} - {1\over a} - {9\over 1-9a+4X+4X^2}
\eqn GammazeroAAone.b. \cr
\AA_0(X)&:={3(X-1)(3-3a+8X+4X^2)\over 8a(a-1)(1-9a+4X+4X^2)}.
\eqn GammazeroAAzero.c.} 
$$
Again, if the integrals are taken over cycles, the right hand side of
\GammazeroInhomPF vanishes, and the independent variable transformation $a\ra
{1\over a}$ gives an equation that is projectively equivalent to the
hypergeometric equation satisfied by $F({1\over 4},{ 1\over 4}, {1\over 2};
{1\over a})$. Choosing $K$ once again to equal a linear combination of the
period integrals as in \lincombK; i.e., expressing $X(a)$ again as an
elliptic function of this argument, this defines a $2$--parameter family of
solutions to the equation obtained by equating the right hand side of
\GammazeroInhomPF to zero. (This again is algebraically related to the
Picard type solutions of $P_{VI}$).  

  A further class of examples is provided by the $4$--vertex cases listed in
Table 2 that correspond to subgroups of $\Gamma$. Up to projective
equivalence, these are $6E\sim \Gamma_0(6)$, $8E'\sim\Gamma_0(8)'\sim
\Gamma_1(4)\cap\Gamma(2)$,  and $9B'\sim\Gamma_0(9)'\sim\Gamma(3)$. (The primes 
$'$ in this notation just denote composition with transformations of the type
$\tau\ra \tau/2$ or $\tau\ra \tau/3$, which do not affect the resulting
Schwarzian equations.)  These are all cases of elliptic pencils corresponding to
Beauville's elliptic surfaces \cite{Be}.  The only case of Beauville's
surfaces that does not appear in Table 2 is the one corresponding
to $\Gamma_1(5)$, which does not give a replicable function since this contains
no $\G_0(N)$ subgroup. Nevertheless, it can be can similarly dealt with.

 As an illustration, here are the details in the case of the Hauptmodul
with automorphism group $\Gamma_0(8)$. The elliptic pencil is defined by
$$
x^3+2xy + a(x^2-y^2)+x=0.  \eq..
$$
In this case, the  elliptic integrals of first and second kinds are
$$
K= \int_\infty^{(X,Y)} {dx\over x-ay}\qquad 
 L= \int_\infty^{(X,Y)} {x dx\over x-ay}. \eq..
$$
The inhomogeneous Gauss-Manin system is
$$
\eqalignno{
K' +{aK+L\over a^2-1} &=
{(1+a^2)X+2aX^2\over a(a^2-1)(X-aY)} + {X'\over X-a Y} \eqn.a.\cr
L' -{aK+L\over a(a^2-1)} &= -{2aX +(1+a^2)X^2\over
a(a^2-1)(X-aY)} + {X X'\over X-aY}, \eqn.b.}
$$
and the inhomogeneous Picard--Fuchs equation is
$$
K''+{(3a^2-1)\over a(a^2-1)}K' +{1\over (a^2-1)}K= 
{1\over X-a Y} \left( X'' -\AA_2(X)X'^2-\AA_1(X)X' -\AA_0(X)\right), \eq..
$$
where
$$
\eqalignno{
\AA_2(X)&:={a+2(1+a^2)X+3aX^2\over 2(aX^3+(1+a^2)X^2+aX)}
\eqn Gammazero_eightAAtwo.a. \cr
\AA_1(X)&:=-{2a^3-(1-4a^2-a^4)X+2a^3X^2\over a(a^2-1)(aX^2+(1+a^2)X+a)}
\eqn Gammazero_eightAAone.b. \cr
\AA_0(X)&:={X(X^2-1)\over 2a(aX^2+(1+a^2)X+a)}.\eqn Gammazero_eightAAzero.a.}
$$
The associated Fuchsian equation for this case is the corresponding 
homogeneous Picard--Fuchs equation, and the result of choosing the
argument of the elliptic function defining $X(a)$ as in \lincombK again defines a
Picard type solution of an equation algebraically related to $P_{VI}$.

   We conclude with the following remarks. 
 \item {1.} The nonlinear monodromy \cite{Du, M} of the Picard--type
solutions of these $P_{VI}$--like equations coincides in each case  with the
monodromy of the associated homogeneous Picard--Fuchs equation, since it just
involves a linear transformation of the coefficients in the linear
combination \lincombK, giving the argument of the elliptic function that defines
the solution $X(a)$.
 
\item {2.}  The Fuchsian and associated Schwarzian equations governing the other
cases of Hauptmoduls, which do not involve subgroups of $\G$,  may also be
derived from Picard--Fuchs equations, since they are obtained by taking
extensions of the automorphism groups by Atkin-Lehner involutions \cite{CN,
FMN}. Instead of considering families of elliptic curves, however, one must
consider families of {\it surfaces} obtained essentially by taking the
topological products of the pairs curves involved. (Further details will be
provided in \cite{H}.)
 
\item {3.} The various $P_{VI}$--like equations that arise are all related by the
same algebraic transformations that connect the corresponding  homogeneous
Picard--Fuchs (and Schwarzian) equations \cite{HM}.

  From the viewpoint of generalizations and applications of integrable dynamical
systems involving such modular functions, it seems natural to try to extend these
considerations to families of higher genus curves, and also to determine
whether analogs of the Chazy solutions exist, which might provide new cases of
Frobenius manifolds.

\bigskip \bigskip 
\noindent\eightpoint{ {\it Acknowledgements.}
I would like to thank J. McKay, J. Hurtubise and C. Doran for helpful
discussions relating to this material. The results quoted in Section
2, in particular the contents of Tables 1 and 2, are taken from
ref.~\cite{HM}, where the associated dynamical systems and the algebraic
relations between the cases are developed in greater detail.  This research was
supported in part by the Natural Sciences and Engineering Research Council of
Canada and the Fonds FCAR du Qu\'ebec.} 

\bigskip \bigskip
  \centerline{\bf References}
\medskip  
 {\eightpoint
\item{\bf [AH]} Atiyah, M.F., and Hitchin, N.J., {\it The Geometry and Dynamics
of  Magnetic Monopoles}, Princeton University Press, Princeton (1988).
\item{\bf [Be]} Beauville, A. ``Les familles stables de courbes elliptiques
sur $\bfP^1$ admettant quatre fibres singuli\'eres, {\it
C.~R.~Acad.~Sci.~Paris\/} {\bf 294}, 6557--660 (1881).
 \item{\bf[Br]} Brioschi, M., , ``Sur un syst\`eme d'\'equations 
diff\'erentielles'', {\it C.~R.~Acad.~Sci.~Paris\/} {\bf 92}, 1389--1393
(1881).
 \item{\bf[C]} Chazy, J., ``Sur les \'equations diff\'erentielles dont 
l'int\'egrale g\'en\'erale poss\`ede une coupure essentielle mobile'',
{\it C.~R.~Acad.~Sc.~Paris}, {\bf 150}, 456--458 (1910). 
 \item{\bf[CAC]} Chakravarty, S., Ablowitz, M.J., and Clarkson, P.A.,
``Reductions of Self-Dual Yang-Mills Fields and Classical Systems'', {\it Phys.
Rev. Let.}  {\bf 65}, 1085--1087 (1990). 
 \item{\bf[CN]} Conway, J. and Norton, S.~P., ``Monstrous moonshine'', 
{\it Bull.~Lond.~Math.~Soc.} {\bf 11}, 308--339 (1979).
\item{\bf [Du]} Dubrovin, B.A., ``Geometry of $2D$ topological field
theories'', Lecture Notes in Math. {\bf 1620}, Springer-Verlag, Berlin,
Heidelberg,  New York (1996). 
\item{\bf [FMN]} Ford, D., McKay, J.,  and Norton, S., ``More on replicable 
functions'' {\it Comm.~in Algebra} {\bf 22}, 5175--5193 (1994).
\item{\bf [F]} Fuchs, R., ``Uber lineare homogene Differentialgleichungen
zweiter Ordnung mit im endlich gelegene wesentlich singul\"aren Stellen.'' {\it 
Math.~Ann.} {\bf 63}, 301--321 (1907).
\item{\bf [GP]} Gibbons, G.W., and Pope, C.N., ``The Positive Action
Conjecture and Asymptotically Euclidean Metrics in Quantum Gravity'',  {\it
Commun.~Math.~Phys.} {\bf 66}, 267--290 (1979). 
\item{\bf [GS]} Gerretson J. and  Sansone, G., {\it Lectures on the Theory of
Functions of a Complex Variable. II. Geometric Theory}. Walters--Noordhoff,
Gr\"oningen (1969).
\item{\bf [H]} Harnad, J., ``Picard--Fuchs equations for elliptic families
and Schwarzian equations for Hauptmoduls'', (CRM preprint (1999), in
preparation).
\item{\bf [Ha]} Halphen, G.-H., ``Sur des fonctions qui proviennent de 
 l'\'equation de Gauss'',  {\it C. R. Acad. Sci. Paris} {\bf 92},
856--858 (1881); ``Sur un syst\`eme d'\'equations   diff\'erentielles'',
{\it ibid.} {\bf 92}, 1101--1103 (1881);
``Sur certains syst\`emes d'\'equations diff\'erentielles'',
{\it ibid.} {\bf 92}, 1404--1406 (1881).
\item{\bf [Hi]} Hitchin, N, ``Twistor Spaces, Einstein metrics and
isomondromic deformations'', {\it J.~Diff.~Geom.} {\bf 42}, 30--112 (1995).
\item{\bf [HM]} Harnad, J. and McKay, J., ``Modular Solutions to Equations
of Generalized Halphen Type'', preprint CRM-2536 (1998),  solv-int/98054006
\item{\bf [Oh]} Ohyama, Yousuke, ``Systems of nonlinear differential 
equations related to second order linear equations'', {\it Osaka J.~Math.}
{\bf 33}, 927--949 (1996);``Differential equations for modular forms with 
level three'', Osaka Univ. ~preprint (1997).
\item{\bf [M]} Mazzocco, M., ``Picard and Chazy Solutions to the Painlev\'e VI
Equation'', preprint SISSA 89/98 (1998).
\item{\bf [Ma]} Manin, Yu., ``Sixth Painlev\'e equations, universal elliptic
curve and mirror of $\bfP^2$'', Bonn preprint (1998), alg--geom/9605010.
\item{\bf [Ta]} Takeuchi, K., ``Arithmetic triangle groups'', {\it
J.~Math.~Soc.~Japan}  {\bf 29}, 91--106 (1977).
\item{\bf [T]} Tod, K.P., ``Self--dual Einstein metrics from the Painlev\'e VI
equation'', {\it Phys.~Lett.} {\bf A190}, 3--4 (1994).
}
\vfill \eject

\end